\documentclass[12pt,a4paper]{article}
\makeatletter
\def\eqnarray{%
\stepcounter{equation}%
\let\@currentlabel=\theequation
\global\@eqnswtrue
\global\@eqcnt\z@
\tabskip\@centering
\let\\=\@eqncr
$$\halign to \displaywidth\bgroup\@eqnsel\hskip\@centering
$\displaystyle\tabskip\z@{##}$&\global\@eqcnt\@ne
\hfil$\displaystyle{{}##{}}$\hfil
&\global\@eqcnt\tw@$\displaystyle\tabskip\z@{##}$\hfil
\tabskip\@centering&\llap{##}\tabskip\z@\cr}
\makeatother
\newcommand{\kansu}[2]{{{#1}\!\left({#2}\right)}}
\newcommand{\ket}[1]{{\vert{#1}\rangle}}
\newcommand{\bra}[1]{{\langle{#1}\vert}}

\newcommand{\calh}{{\cal H}}
\newcommand{\calm}{{\cal M}}
\newcommand{\cala}{{\cal A}}
\newcommand{\calf}{{\cal F}}

\newcommand{\fukuso}{{\mathbf C}}
\newcommand{\futon}{{\bf N}}

\newcommand{\stm}{{St_m}}
\newcommand{\grm}{{Gr_m}}
\newcommand{\eem}{{E_m}}

\newcommand{\xizeta}{{\vert\xi\vert}}
\newcommand{\xis}[1]{{\xi_{#1}}}
\newcommand{\xiszeta}[1]{{\vert\xi_{#1}\vert}}
\newcommand{\etas}[1]{{\eta_{#1}}}
\newcommand{\etaszeta}[1]{{\vert\eta_{#1}\vert}}
\newcommand{\zezeta}{{\vert\zeta\vert}}
\newcommand{\zetas}[1]{{\zeta_{#1}}}
\newcommand{\zetaszeta}[1]{{\vert\zeta_{#1}\vert}}
\newcommand{\kappas}[1]{{\kappa_{#1}}}
\newcommand{\kappaszeta}[1]{{\vert\kappa_{#1}\vert}}
\newcommand{\lam}{{\bf \lambda}}
\newcommand{\slam}{{\bf \lambda_0}}

\begin{document}

\title{\sl More on Optical Holonomic Quantum Computer}
\author{
  Kazuyuki FUJII
  \thanks{E-mail address : fujii@math.yokohama-cu.ac.jp }\\
  Department of Mathematical Sciences\\
  Yokohama City University\\
  Yokohama, 236-0027\\
  Japan
  }
\date{}
\maketitle\thispagestyle{empty}
%
%
%  gaiyou
%
%
\begin{abstract}
   We in this paper consider a further generalization of the (optical)
  holonomic quantum computation proposed by Zanardi and Rasetti   
  (quant--ph 9904011), and reinforced by Fujii (quant--ph 9910069) 
  and Pachos and Chountasis (quant--ph 9912093).

  We construct a quantum computational bundle on some parameter space,  
  and calculate non-abelian Berry connections and curvatures explicitly
  in the special cases.  

  Our main tool is unitary coherent operators based on Lie algebras
  $su(n+1)$ and $su(n,1)$, where the case of $n = 1$ is the 
  previous one.
\end{abstract}

\newpage

%
%
%     Honbun
%
%

\section{Introduction}

This paper is a continuation of Fujii \cite{KF2} and Pachos and Chountasis
\cite{PC} and the aim is to give a mathematical reinforcement to 
\cite{ZR}. 

Quantum Computer is a very attractive and challenging object for New 
Science.

After the breakthrough by P. Shor \cite{PS} there has been remarkable
progress in Quantum Computation (or Computer)(QC briefly).
This discovery had a great influence on scientists. This drived not only 
theoreticians to finding other quantum algorithms, but also 
experimentalists to building quantum computers. See \cite{LPS} in outline.
 \cite{AS} and \cite{RP} are also very useful for non-experts.

On the other hand, Gauge Theories are widely recognized as the basis in 
quantum field theories.
Therefore it is very natural to intend to include  gauge theories 
in QC $\cdots$ a construction of ``gauge theoretical'' quantum computation
or of ``geometric'' quantum computation in our terminology. 
The merit of geometric method of QC is strong against the influence 
from the environment. See \cite{JVEC}. 

Zanardi and Rasetti in \cite{ZR} and  \cite{PZR} proposed such an idea 
using non-abelian Berry phase (quantum holonomy), see also \cite{JP} 
and  \cite{AYK}.
In their model a Hamiltonian (including some parameters) must be
degenerated because an adiabatic connection is introduced using
this degeneracy \cite{SW}. 
In other words, a quantum computational bundle on some parameter space 
(see \cite{ZR}) is introduced due to this degeneracy.

They gave a simple but interesting example to explain their idea. We 
believe that this example will become important in the near future. 
Therefore we treated it once more and gave an explicit form to the 
non-abelian Berry connections and curvatures, see Fujii \cite{KF1} and 
Pachos and Chountasis \cite{PC}.

In \cite{KF1} a non-abelian Berry connection and curvature was 
calculated by making use of the product of unitary coherent operators 
based on Lie algebras $\fukuso$ and $su(1,1)$. On the other hand in 
\cite{PC} an another non-abelian Berry connection and curvature was 
calculated by making use of the product of unitary coherent operators 
based on Lie algebras $su(2)$ and $su(1,1)$. See also \cite{KF2}.

We want to generalize the results above. Namely we want to replace 
Lie algebras $su(2)$ and $su(1,1)$ with bigger Lie algebras $su(n+1)$ 
and $su(n,1)$ for $n \geq 1$. Fortunately we have many studies of 
coherent states based on $su(n+1)$ and $su(n,1)$, see, for examples, 
\cite{FKSF1},\cite{FKSF2} and \cite{FKS}.

In conclusion it is not easy for us to generalize the result in \cite{KF1}
up to this time because we meet some difficulty. But it is, in principle, 
possible to generalize the one in \cite{PC} in spite of hard calculation. 
We show this point in this paper. We list full calculations in the case of 
$n = 2$ and leave the remaining cases to interested readers.

It is not easy to predict the future of geometric quantum computation.
However it is an arena worth challenging for mathematical physicists.

\vspace{5mm}
\section{Mathematical Foundation of Quantum Holonomy}

We start with mathematical preliminaries.
Let $\calh$ be a separable Hilbert space over $\fukuso$.
For $m\in{\bf N}$, we set
\begin{equation}
  \label{eq:stmh}
  \kansu{\stm}{\cal H}
  \equiv
  \left\{
    V=\left( v_1,\cdots,v_m\right)
    \in
    \calh\times\cdots\times\calh\vert V^\dagger V=1_m\right\}\ ,
\end{equation}
where $1_m$ is a unit matrix in $\kansu{M}{m,\fukuso}$.
This is called a (universal) Stiefel manifold.
Note that the unitary group $U(m)$ acts on $\kansu{\stm}{\calh}$
from the right:
\begin{equation}
  \label{eq:stmsha}
  \kansu{\stm}{\calh}\times\kansu{U}{m}
  \rightarrow
  \kansu{\stm}{\calh}: \left( V,a\right)\mapsto Va.
\end{equation}
Next we define a (universal) Grassmann manifold
\begin{equation}
  \kansu{\grm}{\calh}
  \equiv
  \left\{
    X\in\kansu{M}{\calh}\vert
    X^2=X, X^\dagger=X\  \mathrm{and}\  \mathrm{tr}X=m\right\}\ ,
\end{equation}
where $M(\calh)$ denotes a space of all bounded linear operators on $\calh$.
Then we have a projection
\begin{equation}
  \label{eq:piteigi}
  \pi : \kansu{\stm}{\calh}\rightarrow\kansu{\grm}{\calh}\ ,
  \quad \kansu{\pi}{V}\equiv VV^\dagger\ ,
\end{equation}
compatible with the action (\ref{eq:stmsha}) 
($\kansu{\pi}{Va}=Va(Va)^\dagger=Vaa^\dagger V^\dagger=VV^\dagger=
\kansu{\pi}{V}$).

Now the set
\begin{equation}
  \label{eq:principal}
  \left\{
    \kansu{U}{m}, \kansu{\stm}{\calh}, \pi, \kansu{\grm}{\calh}
  \right\}\ ,
\end{equation}
is called a (universal) principal $U(m)$ bundle, 
see \cite{MN} and \cite{KF3}.\quad We set
\begin{equation}
  \label{eq:emh}
  \kansu{\eem}{\cal H}
  \equiv
  \left\{
    \left(X,v\right)
    \in
    \kansu{\grm}{\calh}\times\calh \vert Xv=v \right\}\ .
\end{equation}
Then we have also a projection 
\begin{equation}
  \label{eq:piemgrm}
  \pi : \kansu{\eem}{\calh}\rightarrow\kansu{\grm}{\calh}\ ,
  \quad \kansu{\pi}{\left(X,v\right)}\equiv X\ .
\end{equation}
The set
\begin{equation}
  \label{eq:universal}
  \left\{
    \fukuso^m, \kansu{\eem}{\calh}, \pi, \kansu{\grm}{\calh}
  \right\}\ ,
\end{equation}
is called a (universal) $m$-th vector bundle. This vector bundle is 
one associated with the principal $U(m)$ bundle (\ref{eq:principal})
.

Next let $M$ be a finite or infinite dimensional differentiable manifold 
and the map $P : M \rightarrow \kansu{\grm}{\calh}$ be given (called a 
projector). Using this $P$ we can make 
the bundles (\ref{eq:principal}) and (\ref{eq:universal}) pullback 
over $M$ :
\begin{eqnarray}
  \label{eq:hikimodoshi1}
  &&\left\{\kansu{U}{m},\widetilde{St}, \pi_{\widetilde{St}}, M\right\}
  \equiv
  P^*\left\{\kansu{U}{m}, \kansu{\stm}{\calh}, \pi, 
  \kansu{\grm}{\calh}\right\}
  \ , \\
  \label{eq:hikimodoshi2}
  &&\left\{\fukuso^m,\widetilde{E}, \pi_{\widetilde{E}}, M\right\}
  \equiv
  P^*\left\{\fukuso^m, \kansu{\eem}{\calh}, \pi, \kansu{\grm}{\calh}\right\}
  \ ,
\end{eqnarray}
see \cite{MN}. (\ref{eq:hikimodoshi2}) is of course a vector bundle 
associated with (\ref{eq:hikimodoshi1}).

Let $\calm$ be a parameter space and we denote by $\lam$ its element. 
Let $\slam$ be a fixed reference point of $\calm$. Let $H_\lam$ be 
a family of Hamiltonians parametrized by $\calm$ which act on a Fock space 
$\calh$. We set $H_0$ = $H_\slam$ for simplicity and assume that this has 
a $m$-fold degenerate vacuum :
\begin{equation}
  H_{0}v_j = \mathbf{0},\quad j = 1 \sim m. 
\end{equation}
These $v_j$'s form a $m$-dimensional vector space. We may assume that 
$\langle v_{i}\vert v_{j}\rangle = \delta_{ij}$. Then $\left(v_1,\cdots,v_m
\right) \in \kansu{\stm}{\calh}$ and 
\[
  F_0 \equiv \left\{\sum_{j=1}^{m}x_{j}v_{j}\vert x_j \in \fukuso \right\} 
  \cong \fukuso^m.
\]
Namely, $F_0$ is a vector space associated with o.n.basis 
$\left(v_1,\cdots,v_m\right)$.

Next we assume for simplicity 
that a family of unitary operators parametrized by $\calm$
\begin{equation}
  \label{eq:ufamily} 
  W : \calm \rightarrow U(\calh),\quad W(\slam) = {\rm id}.
\end{equation}
is given and $H_{\lam}$ above is given by the following isospectral family
\begin{equation}
 H_{\lam} \equiv W(\lam)H_0 W(\lam)^{-1}.
\end{equation}
In this case there is no level crossing of eigenvalues. Making use of 
$W(\lam)$ we can define a projector
\begin{equation}
  \label{eq:pfamily}
 P : \calm \rightarrow \kansu{\grm}{\calh}, \quad 
 P(\lam) \equiv
  W(\lam) \left(\sum^{m}_{j=1}v_{j}v_{j}^{\dagger}\right)W(\lam)^{-1}
\end{equation}
and have the pullback bundles over $\calm$
\begin{equation}
  \label{eq:target}
 \left\{\kansu{U}{m},\widetilde{St}, \pi_{\widetilde{St}}, \calm\right\},\quad 
 \left\{\fukuso^m,\widetilde{E}, \pi_{\widetilde{E}}, \calm\right\}.
\end{equation}

For the later we set
\begin{equation}
  \label{eq:vacuum}
 \ket{vac} = \left(v_1,\cdots,v_m\right).
\end{equation}
In this case a canonical connection form $\cala$ of 
$\left\{\kansu{U}{m},\widetilde{St}, \pi_{\widetilde{St}}, \calm\right\}$ is 
given by 
\begin{equation}
  \label{eq:cform}
 \cala = \bra{vac}W(\lam)^{-1}d W(\lam)\ket{vac},
\end{equation}
where $d$ is a differential form on $\calm$, and its curvature form by
\begin{equation}
  \label{eq:curvature}
  \calf \equiv d\cala+\cala\wedge\cala.
\end{equation}
On the other hand the global form of the curvature is given by 
$PdP\wedge dP$, which is related to (\ref{eq:curvature}) by
\begin{equation}
  \label{eq:projector}
  PdP\wedge dP  = W\calf W^{-1} = 
  W\left(d\cala+\cala\wedge\cala\right) W^{-1}\ ,
\end{equation}
see \cite{SW}, \cite{MN} and \cite{KF1}.

Let $\gamma$ be a loop in $\calm$ at $\slam$., $\gamma : [0,1] 
\rightarrow \calm, \gamma(0) = \gamma(1)$. For this $\gamma$ a holonomy 
operator $\Gamma_{\cala}$ is defined :
\begin{equation}
  \label{eq:holonomy}
  \Gamma_{\cala}(\gamma) = {\cal P}exp\left\{\oint_{\gamma}\cala\right\} 
  \in \kansu{U}{m},
\end{equation}
where ${\cal P}$ means path-ordered. This acts on the fiber $F_0$ at 
$\slam$ of the vector bundle 
$\left\{\fukuso^m,\widetilde{E}, \pi_{\widetilde{E}}, M\right\}$ as follows :
${\textbf x} \rightarrow \Gamma_{\cala}(\gamma){\textbf x}$.\quad 
The holonomy group $Hol(\cala)$ is in general subgroup of $\kansu{U}{m}$ 
. In the case of $Hol(\cala) = \kansu{U}{m}$,   $\cala$ is called 
irreducible, see \cite{MN}.

In the Holonomic Quantum Computer we take  
\begin{eqnarray}
  \label{eq:information}
  &&{\rm Encoding\ of\ Information} \Longrightarrow {\textbf x} \in F_0 , 
  \nonumber \\
  &&{\rm Processing\ of\ Information} \Longrightarrow \Gamma_{\cala}(\gamma) : 
  {\textbf x} \rightarrow \Gamma_{\cala}(\gamma){\textbf x}.
\end{eqnarray}

%\vspace{5mm}
\section{Unitary Coherent Operators based on $su(n+1)$ and $su(n,1)$}

We apply the results of last section to Quantum Optics and discuss about 
(optical) Holonomic Quantum Computation proposed by \cite{ZR} and 
\cite{PC}.

Let $a(a^\dagger)$ be the annihilation (creation) operator of the harmonic 
oscillator.
If we set $N\equiv a^\dagger a$ (:\ number operator), then
\begin{equation}
  [N,a^\dagger]=a^\dagger\ ,\
  [N,a]=-a\ ,\
  [a,a^\dagger]=1\ .
\end{equation}
Let $\calh$ be a Fock space generated by $a$ and $a^\dagger$, and
$\{\ket{n}\vert n\in\futon\cup\{0\}\}$ be its basis.
The actions of $a$ and $a^\dagger$ on $\calh$ are given by
\begin{equation}
  \label{eq:shoukou}
  a\ket{n} = \sqrt{n}\ket{n-1}\ ,\
  a^\dagger\ket{n} = \sqrt{n+1}\ket{n+1}\ ,
\end{equation}
where $\ket{0}$ is a vacuum ($a\ket{0}=0$).

Next we consider the system of $n+1$--harmonic oscillators. For $1\leq j 
\leq n+1$ we set
\begin{eqnarray}
  \label{eq:twosystem}
  a_j &=& 1 \otimes \cdots \otimes 1 \otimes a \otimes 1 \otimes \cdots 
        \otimes 1 \ ({\rm j-position}),
  \nonumber \\
  {a_j}^{\dagger} &=& 
   1 \otimes \cdots \otimes 1 \otimes a^{\dagger} \otimes 1 \otimes 
   \cdots \otimes 1 \ ({\rm j-position}),
\end{eqnarray}
then it is easy to see 
\begin{equation}
  \label{eq:relations}
 [a_i, a_j] = [a_i^{\dagger}, a_j^{\dagger}] = 0,\ 
 [a_i, a_j^{\dagger}] = \delta_{ij}, \quad i, j = 1, 2, \cdots, n+1. 
\end{equation}
We also denote by $N_{j} = a_j^{\dagger}a_j$ its number operators.

The Fock space $\calh^{(n+1)}$ fot the system of $n+1$--harmonic oscillators 
is the $n+1$--tensor product $\calh^{(n+1)}=\calh \otimes \cdots \otimes 
\calh$, and  
each $a_j\ {\rm and}\ {a_j}^{\dagger}$ acts on $j$--component of 
$\calh^{(n+1)}$ like (\ref{eq:shoukou}).

Now since we want to consider coherent states based on Lie algebras $su(n+1)$ 
and $su(n,1)$, we make use of Schwinger's boson method, see \cite{FKSF1}, 
\cite{FKSF2}. 

\subsection{$su(n+1)$}

If we set
\begin{equation}
  \label{eq:c-generators}
   E_{\alpha\beta}= {a_\alpha}^{\dagger}a_\beta,\quad 
                     1\leq \alpha,\ \beta \leq n+1 
\end{equation}
then from (\ref{eq:relations}) we find 
\begin{equation}
  \label{eq:c-relations}
    [E_{\alpha\beta}, E_{\gamma\delta}] = 
    E_{\alpha\delta}{\mathbf \delta}_{\beta\gamma} - 
    {\mathbf \delta}_{\delta\alpha}E_{\gamma\beta},
\end{equation}
where ${\mathbf \delta} = {\rm diag.}(1, \cdots, 1)$. That is,  
$\left\{E_{\alpha\beta} \vert\  1\leq \alpha,\ \beta \leq n+1 \right\}$ 
is a generator of Lie algebra $u(n+1)$. 
Then a set of generator $\left\{
 E_{j,n+1} \vert 1 \leq j \leq n \right\}$ plays an important role.
For $1 \leq j \leq n$ we set
\begin{equation}
%  \label{eq:J-daisu}
    {J^j}_+ = a_j^{\dagger}a_{n+1},\ {J^j}_- = {a_{n+1}}^{\dagger}a_j,\ 
    {J^j}_3 = {1\over2}\left(a_j^{\dagger}a_j - 
              {a_{n+1}}^{\dagger}a_{n+1}\right), 
\end{equation}
then we have 
\begin{equation}
%  \label{eq:j-relation}
     [{J^j}_3,{J^j}_+ ] = {J^j}_+ ,\
     [{J^j}_3,{J^j}_- ] = -{J^j}_- ,\
     [{J^j}_+, {J^j}_- ] = 2 {J^j}_3.
\end{equation}
Namely, $\left\{{J^j}_+,{J^j}_-, {J^j}_3 \right\}$ forms $su(2)$--algebra. 
From this we can construct unitary coherent operators based on 
$su(2)$--algebra : For $1 \leq j \leq n$
\begin{equation}
  U_j(\xis{j}) = exp{\left(\xis{j} a_j^{\dagger}a_{n+1} - 
                {\bar \xis{j}}{a_{n+1}}^{\dagger}a_j \right)}. 
\end{equation}
The disentangling formula for this operator is 
\begin{equation}
   U_j(\xis{j}) = e^{\etas{j} a_j^{\dagger}a_{n+1}}
 e^{{\rm log}\left(1 + {\etaszeta{j}}^{2}\right)
    {1\over2}\left(a_j^{\dagger}a_j - a_{n+1}^{\dagger}a_{n+1}\right)}
           e^{-{\bar \eta}_j{a_{n+1}}^{\dagger}a_j}, \quad 
   \etas{j} = \frac{\xis{j}\tan\left(\xiszeta{j}\right)}{\xiszeta{j}}.
\end{equation}
see \cite{KF2}. 
Therefore combining these operators  
we define a unitary coherent operator based on $su(n+1)$--algebra :

\noindent{\bfseries Definition 1}\quad For $\vec{\xi} = \left(\xis{1},
\xis{2}, \cdots, \xis{n}\right)$ we define 
\begin{equation}
  \label{eq:c-ucoherent}
   U\left(\vec{\xi}\ \right) \equiv \prod_{j=1}^{n} U_j(\xis{j})\quad 
   ({\rm in\ this\ order}). 
\end{equation}
Fot simplicity we also set $U\left(\vec{\xi}\ \right)= 
\prod_{j=1}^{n} U_j$.
For the latter convenience let us calculate $U\left(\vec{\xi}\ \right)^{-1}
\frac{\partial}{\partial\xis{j}}U\left(\vec{\xi}\ \right)$. It is easy to 
see
\[
  U\left(\vec{\xi}\ \right)^{-1}\frac{\partial}{\partial\xis{j}}
  U\left(\vec{\xi}\ \right) 
  = U_{n}^{-1}\cdots U_{j+1}^{-1}\left(U_{j}^{-1}
    \frac{\partial}{\partial\xis{j}}U_{j}\right)U_{j+1}\cdots U_{n}. 
\]
On the other hand  we have already calculated $U_{j}^{-1}\frac{\partial}
{\partial\xis{j}}U_{j}$ :

\noindent{\bfseries Lemma 2}\quad We have
\begin{eqnarray}
 U_{j}^{-1}\frac{\partial}{\partial\xis{j}}U_{j} = &&
  {1\over2}\left(1+{\sin(2\xiszeta{j})\over2\xiszeta{j}}\right)
   a_j^{\dagger}a_{n+1} \nonumber \\
   &+& 
  {{\bar \xis{j}}\over2\xiszeta{j}^{2}}\left(1-\cos(2\xiszeta{j})\right)
  {1\over2}\left(a_j^{\dagger}a_j -  a_{n+1}^{\dagger}a_{n+1}\right)
    \nonumber \\  
   &+& 
  {{\bar \xis{j}}^{2}\over2\xiszeta{j}^{2}}
  \left(-1+{\sin(2\xiszeta{j})\over2\xiszeta{j}}\right)
  a_{n+1}^{\dagger}a_j .
\end{eqnarray}
From this we easily obtain 
\begin{eqnarray}
  U\left(\vec{\xi}\ \right)^{-1}\frac{\partial}{\partial\xis{j}}
  U\left(\vec{\xi}\ \right)  
  = &&
  {1\over2}\left(1+{\sin(2\xiszeta{j})\over2\xiszeta{j}}\right)
  a_j^{\dagger}U_{n}^{-1}\cdots U_{j+1}^{-1}a_{n+1}U_{j+1}\cdots U_{n}
  \nonumber \\ 
    &+&
  {{\bar \xis{j}}\over2\xiszeta{j}^{2}}\left(1-\cos(2\xiszeta{j})\right)
  {1\over2}\left(a_j^{\dagger}a_j - 
  U_{n}^{-1}\cdots U_{j+1}^{-1}a_{n+1}^{\dagger}a_{n+1}U_{j+1}\cdots U_{n}
  \right)  \nonumber \\ 
    &+&
  {{\bar \xis{j}}^{2}\over2\xiszeta{j}^{2}}
  \left(-1+{\sin(2\xiszeta{j})\over2\xiszeta{j}}\right)
  U_{n}^{-1}\cdots U_{j+1}^{-1}a_{n+1}^{\dagger}U_{j+1}\cdots U_{n}a_j   
\end{eqnarray}
Therefore we have only to calculate the term $
   U_{n}^{-1}\cdots U_{j+1}^{-1}a_{n+1}U_{j+1}\cdots U_{n} 
$.

\noindent{\bfseries Lemma 3}\quad We have
\begin{eqnarray}
   U_{n}^{-1}\cdots U_{j+1}^{-1}&a_{n+1}&U_{j+1}\cdots U_{n} 
   = c_{n,j+1}a_{n+1} - \sum_{l=j+1}^{n}d_{l,j+1}a_{l},
  \quad{\rm where}    \nonumber \\ 
 c_{n,j+1}&\equiv& \cos(\xiszeta{n})\cos(\xiszeta{n-1})\cdots 
                   \cos(\xiszeta{j+1}),
   \nonumber  \\ 
 d_{l,j+1}&\equiv& {{\bar \xis{l}}\sin(\xiszeta{l})\over\xiszeta{l}}
              \cos(\xiszeta{l-1})\cos(\xiszeta{l-2})\cdots \cos(\xiszeta{j+1}),
    \nonumber \\ 
             && {\rm for}\quad j+1 \leq l \leq n.
\end{eqnarray}
Fron these facts  we obtain

\noindent{\bfseries Proposition 4}\quad {\rm for} $1 \leq j \leq n$
\begin{eqnarray}
 &&U\left(\vec{\xi}\ \right)^{-1}\frac{\partial}{\partial\xis{j}}
  U\left(\vec{\xi}\ \right)  \nonumber \\ 
 &=& 
  {1\over2}\left(1+{\sin(2\xiszeta{j})\over2\xiszeta{j}}\right)
   \left\{ c_{n,j+1}a_j^{\dagger}a_{n+1}- \sum_{l=j+1}^{n}d_{l,j+1}
   a_j^{\dagger}a_{l} \right\}  \nonumber \\ 
   &&+
  {{\bar \xis{j}}\over2\xiszeta{j}^{2}}\left(1-\cos(2\xiszeta{j})\right) 
  {1\over2}
  \left\{
   a_j^{\dagger}a_j -  c_{n,j+1}^{2}a_{n+1}^{\dagger}a_{n+1}
  + \sum_{l=j+1}^{n}c_{n,j+1}d_{l,j+1}a_{n+1}^{\dagger}a_{l} 
  \right. \nonumber \\ 
   &&\left. + \sum_{l=j+1}^{n}c_{n,j+1}{\bar d}_{l,j+1}
         a_{l}^{\dagger}a_{n+1} 
  - \sum_{l,k=j+1}^{n}{\bar d}_{l,j+1}d_{k,j+1}a_{l}^{\dagger}a_{k} 
  \right\}  \nonumber \\ 
  &&+
  {{\bar \xis{j}}^{2}\over2\xiszeta{j}^{2}}
  \left(-1+{\sin(2\xiszeta{j})\over2\xiszeta{j}}\right)
  \left\{c_{n,j+1}a_{n+1}^{\dagger}a_{j} - \sum_{l=j+1}^{n}
    {\bar d}_{l,j+1}a_{l}^{\dagger}a_{j} \right\}.
\end{eqnarray}

\subsection{$su(n,1)$}

If we set
\begin{eqnarray}
  \label{eq:nc-generators}
   &&E_{\alpha\beta}= {a_\alpha}^{\dagger}a_\beta,\quad 
                       1\leq \alpha,\ \beta \leq n , \nonumber \\
   &&E_{n+1,\alpha}= a_{n+1}a_\alpha,\quad 
   E_{\alpha,n+1}= {a_\alpha}^{\dagger}{a_{n+1}}^{\dagger},\quad 
   E_{n+1,n+1}={a_{n+1}}^{\dagger}a_{n+1}+1,  
\end{eqnarray}
then from (\ref{eq:relations}) we find 
\begin{equation}
  \label{eq:nc-relations}
    [E_{\alpha\beta}, E_{\gamma\delta}] = 
    E_{\alpha\delta}{\mathbf \eta}_{\beta\gamma} - 
    {\mathbf \eta}_{\delta\alpha}E_{\gamma\beta},
\end{equation}
where ${\mathbf \eta} = {\rm diag.}(1, \cdots, 1, -1)$. That is,  
$\left\{E_{\alpha\beta} \vert\  1\leq \alpha,\ \beta \leq n+1 \right\}$ 
is a generator of Lie algebra $u(n,1)$.
Then a set of generator $\left\{
E_{j,n+1} \vert 1 \leq j \leq n \right\}$ plays an important role.
For $1 \leq j \leq n$ we set
\begin{equation}
%  \label{eq:K-daisu}
    {K^j}_+ = a_j^{\dagger}{a_{n+1}}^{\dagger},\ 
    {K^j}_- = a_{n+1}a_j,\ 
    {K^j}_3 = {1\over2}\left(a_j^{\dagger}a_j + 
              {a_{n+1}}^{\dagger}a_{n+1} + 1 \right), 
\end{equation}
then we have 
\begin{equation}
%  \label{eq:j-relation}
     [{K^j}_3,{K^j}_+ ] = {K^j}_+ ,\
     [{K^j}_3,{K^j}_- ] = -{K^j}_- ,\
     [{K^j}_+, {K^j}_- ] = - 2 {K^j}_3.
\end{equation}
Namely, $\left\{{K^j}_+,{K^j}_-, {K^j}_3 \right\}$ forms $su(1,1)$--algebra. 
From this we can construct unitary coherent operators based on 
$su(1,1)$--algebra : For $1 \leq j \leq n$
\begin{equation}
  V_j(\zetas{j}) = exp{\left(\zetas{j} a_j^{\dagger}{a_{n+1}}^{\dagger} - 
                {\bar \zetas{j}}a_{n+1}a_j \right)}. 
\end{equation}
The disentangling formula for this operator is 
\begin{equation}
   V_j(\zetas{j}) = e^{\kappas{j} a_j^{\dagger}{a_{n+1}}^{\dagger}}
       e^{{\rm log}\left(1 - {\kappaszeta{j}}^{2}\right)
      {1\over2}\left(a_j^{\dagger}a_j + a_{n+1}^{\dagger}a_{n+1} + 1 \right)}
           e^{-{\bar \kappa}_ja_{n+1}a_j}, \quad 
   \kappas{j} = \frac{\zetas{j}\tanh\left(\zetaszeta{j}\right)}
                     {\zetaszeta{j}}.
\end{equation}
see \cite{KF2}. 
Therefore combining these operators  
we define a unitary coherent operator based on $su(n,1)$--algebra :

\noindent{\bfseries Definition 5}\quad For $\vec{\zeta} = \left(\zetas{1},
\zetas{2}, \cdots, \zetas{n}\right)$ we define  
\begin{equation}
  \label{eq:nc-ucoherent}
   V\left(\vec{\zeta}\ \right) \equiv \prod_{j=1}^{n} V_j(\zetas{j})\quad 
   ({\rm in\ this\ order}). 
\end{equation}
Fot simplicity we also set $V\left(\vec{\zeta}\ \right)= 
\prod_{j=1}^{n} V_j$.
For the latter convenience let us calculate $V\left(\vec{\zeta}\ \right)^{-1}
\frac{\partial}{\partial\zetas{j}}V\left(\vec{\xi}\ \right)$. It is easy to 
see
\[
  V\left(\vec{\zeta}\ \right)^{-1}\frac{\partial}{\partial\zetas{j}}
  V\left(\vec{\zeta}\ \right) 
  = V_{n}^{-1}\cdots V_{j+1}^{-1}\left(V_{j}^{-1}
    \frac{\partial}{\partial\zetas{j}}V_{j}\right)V_{j+1}\cdots V_{n}. 
\]
On the other hand  we have already calculated $V_{j}^{-1}\frac{\partial}
{\partial\zetas{j}}V_{j}$ :

\noindent{\bfseries Lemma 6}\quad We have
\begin{eqnarray}
 V_{j}^{-1}\frac{\partial}{\partial\zetas{j}}V_{j} = &&
  {1\over2}\left(1+{\sinh(2\zetaszeta{j})\over2\zetaszeta{j}}\right)
   a_j^{\dagger}a_{n+1}^{\dagger} \nonumber \\
   &+& 
{{\bar \zetas{j}}\over2\zetaszeta{j}^{2}}\left(-1+\cosh(2\zetaszeta{j})\right)
  {1\over2}\left(a_j^{\dagger}a_j + a_{n+1}^{\dagger}a_{n+1} + 1 \right)
    \nonumber \\  
   &+& 
  {{\bar \zetas{j}}^{2}\over2\zetaszeta{j}^{2}}
  \left(-1+{\sinh(2\zetaszeta{j})\over2\zetaszeta{j}}\right)
  a_{n+1}a_j .
\end{eqnarray}
From this we easily obtain 
\begin{eqnarray}
  V\left(\vec{\zeta}\ \right)^{-1}\frac{\partial}{\partial\zetas{j}}
  V\left(\vec{\zeta}\ \right)  
  &=&
  {1\over2}\left(1+{\sinh(2\zetaszeta{j})\over2\zetaszeta{j}}\right)
 a_j^{\dagger}V_{n}^{-1}\cdots V_{j+1}^{-1}a_{n+1}^{\dagger}V_{j+1}\cdots V_{n}
  \nonumber \\ 
    &+&
 {{\bar \zetas{j}}\over2\zetaszeta{j}^{2}}\left(-1+\cosh(2\zetaszeta{j})\right)
  {1\over2}\left(a_j^{\dagger}a_j + 1 + 
  V_{n}^{-1}\cdots V_{j+1}^{-1}a_{n+1}^{\dagger}a_{n+1}V_{j+1}\cdots V_{n}
  \right)  \nonumber \\ 
    &+&
  {{\bar \zetas{j}}^{2}\over2\zetaszeta{j}^{2}}
  \left(-1+{\sinh(2\zetaszeta{j})\over2\zetaszeta{j}}\right)
  V_{n}^{-1}\cdots V_{j+1}^{-1}a_{n+1}V_{j+1}\cdots V_{n}a_j   
\end{eqnarray}
Therefore we have only to calculate the term $
   V_{n}^{-1}\cdots V_{j+1}^{-1}a_{n+1}^{\dagger}V_{j+1}\cdots V_{n} 
$.

\noindent{\bfseries Lemma 7}\quad We have
\begin{eqnarray}
   V_{n}^{-1}\cdots V_{j+1}^{-1}&a_{n+1}^{\dagger}&V_{j+1}\cdots V_{n} 
   = e_{n,j+1}a_{n+1}^{\dagger} + \sum_{l=j+1}^{n}f_{l,j+1}a_{l},
  \quad{\rm where}    \nonumber \\ 
 e_{n,j+1}&\equiv& \cosh(\zetaszeta{n})\cosh(\zetaszeta{n-1})\cdots 
                   \cosh(\zetaszeta{j+1}),
   \nonumber  \\ 
 f_{l,j+1}&\equiv& 
  {{\bar \zetas{l}}\sinh(\zetaszeta{l})\over\zetaszeta{l}}
   \cosh(\zetaszeta{l-1})\cosh(\zetaszeta{l-2})\cdots \cosh(\zetaszeta{j+1}),
    \nonumber \\ 
             && {\rm for}\quad j+1 \leq l \leq n.
\end{eqnarray}
Fron these facts  we obtain

\noindent{\bfseries Proposition 8}\quad {\rm for} $1 \leq j \leq n$
\begin{eqnarray}
 &&V\left(\vec{\zeta}\ \right)^{-1}\frac{\partial}{\partial\zetas{j}}
  V\left(\vec{\zeta}\ \right)  \nonumber \\ 
 &=& 
  {1\over2}\left(1+{\sinh(2\zetaszeta{j})\over2\zetaszeta{j}}\right)
   \left\{e_{n,j+1}a_j^{\dagger}a_{n+1}^{\dagger}
    + \sum_{l=j+1}^{n}f_{l,j+1}a_j^{\dagger}a_{l} \right\}  \nonumber \\ 
   &&+
{{\bar \zetas{j}}\over2\zetaszeta{j}^{2}}\left(-1+\cosh(2\zetaszeta{j})\right) 
  {1\over2}
  \left\{
  a_j^{\dagger}a_j + e_{n,j+1}^{2}\left(a_{n+1}^{\dagger}a_{n+1}+1\right)
  + \sum_{l=j+1}^{n}e_{n,j+1}{\bar f}_{l,j+1}a_{l}^{\dagger}
    a_{n+1}^{\dagger}   \right. \nonumber \\ 
   &&\left. + \sum_{l=j+1}^{n}e_{n,j+1}f_{l,j+1}a_{n+1}a_{l}
  + \sum_{l,k=j+1}^{n}f_{l,j+1}{\bar f}_{k,j+1}a_{l}^{\dagger}a_{k} 
  \right\}  \nonumber \\ 
  &&+
  {{\bar \zetas{j}}^{2}\over2\zetaszeta{j}^{2}}
  \left(-1+{\sinh(2\zetaszeta{j})\over2\zetaszeta{j}}\right)
  \left\{e_{n,j+1}a_{n+1}a_{j} + \sum_{l=j+1}^{n}
    {\bar f}_{l,j+1}a_{l}^{\dagger}a_{j} \right\}.
\end{eqnarray}

\section{Optical Holonomic Quantum Computer $\cdots$ Generalization}

Let $H_0$ be a Hamiltonian with nonlinear interaction produced by 
a Kerr medium., that is  $H_0 = \hbar {\rm X} N(N-1)$, where X is a 
certain constant, see \cite{PC}. The eigenvectors of $H_0$ corresponding 
to $0$ is $\left\{\ket{0},\ket{1}\right\}$, so its eigenspace is 
${\rm Vect}\left\{\ket{0},\ket{1}\right\} \cong \fukuso^2$. We correspond to 
$0 \rightarrow \ket{0},\ 1 \rightarrow \ket{1}$ for a generator of 
Boolean algebra $\left\{0, 1\right\}$. The space 
${\rm Vect}\left\{\ket{0},\ket{1}\right\}$ is called 1-qubit (quantum bit) 
space, see \cite{AS} or  \cite{RP}. 
Since we are considering the system of $n+1$ particles, the Hamiltonian that 
we treat in the following becomes 
\begin{equation}
  \label{eq:hamiltonian}
  H_0 = \sum_{j=1}^{n+1} \hbar {\rm X} N_{j}(N_{j}-1).
\end{equation}
The $0$--eigenspace of this Hamiltonian becomes therefore 
\begin{equation}
  \label{eq:eigenspace}
   F_0 = {\rm Vect}\left\{\ket{0},\ket{1}\right\}
         \otimes \cdots \otimes
         {\rm Vect}\left\{\ket{0},\ket{1}\right\} 
         \cong \fukuso^2\otimes \cdots \otimes \fukuso^2 
         \cong \fukuso^{2^{n+1}}.
\end{equation}
We denote 
\[
   \langle \alpha_1,\cdots,\alpha_{n+1} \vert 
   \beta_1,\cdots,\beta_{n+1} \rangle = 
   \prod_{j=1}^{n+1}\langle \alpha_j \vert \beta_j \rangle = 
   \prod_{j=1}^{n+1}\delta_{\alpha_j \beta_j}, 
\]
for $\vert \alpha_1,\cdots,\alpha_{n+1}\rangle,\ 
\vert \beta_1,\cdots,\beta_{n+1}\rangle \in  F_0$.  
We order the basis of $F_0$ as 
\begin{eqnarray}
   \ket{0}&=& \vert 0,0,\cdots,0,0 \rangle, \nonumber \\
   \ket{1}&=& \vert 0,0,\cdots,0,1 \rangle, \nonumber \\
   \cdots \nonumber \\
   \ket{2^{n+1}-2} &=& \vert 1,1,\cdots,1,0 \rangle, \nonumber \\
   \ket{2^{n+1}-1} &=& \vert 1,1,\cdots,1,1 \rangle. \nonumber 
\end{eqnarray}
and set 
\begin{equation}
 \label{eq:vacuum2}
  \ket{vac}= \left(\ket{0}, \ket{1}, \cdots, \ket{2^{n+1}-1} \right).
\end{equation}
Namely $m=2^{n+1}-1$ in (\ref{eq:vacuum}).

Here we consider the following isospectral family of $H_0$ above :
\begin{eqnarray}
  \label{eq:generalfamily}
   H_{(\vec{\xi}\ ,\vec{\zeta}\ )} &=& 
    W(\vec{\xi}\ ,\vec{\zeta}\ )H_0 W(\vec{\xi}\ ,\vec{\zeta}\ )^{-1},\\
   W(\vec{\xi}\ ,\vec{\zeta}\ ) &=& U(\vec{\xi}\ )V(\vec{\zeta}\ ) \in 
     U(\calh \otimes \cdots \otimes \calh)\ (n+1-{\rm times}),\quad 
   W(\vec{0}\ ,\vec{0}\ ) = {\rm id}.
\end{eqnarray}
For this system we want to calculate a connection form (\ref{eq:cform})
in the last section. For that we set : for $1 \leq j \leq n$ 
\begin{eqnarray}
  \label{eq:coefficients}
  A_{\xis{j}} &=& \bra{vac}W(\vec{\xi}\ ,\vec{\zeta}\ )^{-1}
              \frac{\partial}{\partial \xis{j}}
            W(\vec{\xi}\ ,\vec{\zeta}\ )\ket{vac},   \nonumber \\
  A_{\zetas{j}} &=& \bra{vac}W(\vec{\xi}\ ,\vec{\zeta}\ )^{-1}
            \frac{\partial}{\partial \zetas{j}}
            W(\vec{\xi}\ ,\vec{\zeta}\ )\ket{vac}.
\end{eqnarray}
Here we note   
\begin{eqnarray}
  \label{eq:cartan1}
 W(\vec{\xi}\ ,\vec{\zeta}\ )^{-1}
\frac{\partial}{\partial \xis{j}}
 W(\vec{\xi}\ ,\vec{\zeta}\ ) &=& 
 V(\vec{\zeta}\ )^{-1}\left\{U(\vec{\xi}\ )^{-1}
 \frac{\partial}{\partial \xis{j}}
 U(\vec{\xi}\ )\right\}V(\vec{\zeta}\ ), 
 \\
  \label{eq:cartan2}
 W(\vec{\xi}\ ,\vec{\zeta}\ )^{-1}
\frac{\partial}{\partial \zetas{j}}W(\vec{\xi}\ ,\vec{\zeta}\ ) &=& 
 V(\vec{\zeta}\ )^{-1}\frac{\partial}{\partial \zetas{j}}V(\vec{\zeta}\ )  
 \nonumber.
\end{eqnarray}
On the other hand in Proposition 4 and Proposition 8 we have already 
calculated  \\  \noindent
$U(\vec{\xi}\ )^{-1}\frac{\partial}{\partial \xis{j}}U(\vec{\xi}\ )$ and 
$V(\vec{\zeta}\ )^{-1}\frac{\partial}{\partial \zetas{j}}V(\vec{\zeta}\ )$. 

From Proposition 4 we must calculate $V^{-1}a_{\alpha}^{\dagger}a_{\beta}
V = \left(V^{-1}a_{\alpha}V\right)^{\dagger}\left(V^{-1}a_{\beta}V\right),
\ {\rm where}\  V = V_1V_2\cdots V_n$. Therefore let us calculate 
$V^{-1}a_{\alpha}V \quad {\rm for}\ 1 \leq \alpha \leq n+1$. But 
remarking that $V_j a_k = a_k V_j \quad {\rm for}\ 1 \leq k \leq j-1$ 
because 
$V_j \equiv 
  V_j(\zetas{j}) = {\rm exp}{\left(\zetas{j} a_j^{\dagger}{a_{n+1}}^{\dagger} 
                    - {\bar \zetas{j}}a_{n+1}a_j \right)}
$ 
we must calculate 
\begin{eqnarray}
%\label{eq:adjoint-j}
   V^{-1}a_j V &=& V_n^{-1}\cdots V_j^{-1}a_jV_j\cdots V_n 
   \qquad {\rm for}\ 1 \leq j \leq n , \nonumber \\
%\label{eq:adjoint-n+1}
   V^{-1}a_{n+1}V &=& V_n^{-1}\cdots V_1^{-1}a_{n+1}V_1\cdots V_n.
                                     \nonumber  
\end{eqnarray}
To calculate these is not so difficult.  The result is

\noindent{\bfseries Lemma 9}\quad 
\begin{eqnarray}
\label{eq:V-j}
 V^{-1}a_j V &=& 
\cosh(\zetaszeta{j})a_{j} + 
\frac{\zetas{j}\sinh(\zetaszeta{j})}{\zetaszeta{j}}
\left\{ \sum_{l=j+1}^{n}\prod_{k=j+1}^{l-1}
\cosh(\zetaszeta{k})\frac{{\bar \zetas{l}}\sinh(\zetaszeta{l})}{\zetaszeta{l}}
a_{l} \right. \nonumber \\
&+& 
\left. \prod_{k=j+1}^{n}\cosh(\zetaszeta{k})a_{n+1}^{\dagger} \right\} 
\qquad {\rm for}\ 1 \leq j \leq n \ ,  \\
\label{eq:V-n+1}
V^{-1}a_{n+1}V &=& 
\sum_{k=1}^{n}\left\{ \prod_{l=1}^{k-1}\cosh(\zetaszeta{l})\right\}
\frac{\zetas{k}\sinh(\zetaszeta{k})}{\zetaszeta{k}}a_{k}^{\dagger} + 
\prod_{l=1}^{n}\cosh(\zetaszeta{l})a_{n+1}.
\end{eqnarray}
Here we should understand that $\prod_{k}^{l}(\cdots) = 1$ if $l < k$.

Using this we can calculate $V^{-1}a_{\alpha}^{\dagger}a_{\beta}V$ and next 
calculate (\ref{eq:cartan1}) in principle. 
But it is not easy for us to obtain a compact form for this 
up to this time. Therefore let us restrict to some special cases ($n = 1, 2$) 
and obtain complete forms.

\subsection{Example $\cdots \ n=1$}

In this case we can obtain the connection form in a complete manner. 
See \cite{KF2} and also \cite{PC}. For simplicity we set $\xis{1}=\xi$ 
and $\zetas{1}=\zeta$. The result is

\noindent{\bfseries Lemma 10}\quad we have
\begin{eqnarray}
 &&W^{-1}\frac{\partial}{\partial \xi}W   \nonumber \\
 &=& 
 {1\over2}\left(1+{\sin(2\xizeta)\over2\xizeta}\right)
 \left\{
 \cosh(2\zezeta)a_1^{\dagger}a_2 + 
 {\zeta\sinh(2\zezeta)\over2\zezeta}\left(a_1^{\dagger}\right)^2 +
 {{\bar \zeta}\sinh(2\zezeta)\over2\zezeta}\left(a_2\right)^2 
 \right\} \nonumber \\
 &+& {{\bar \xi}\over2\xizeta^{2}}\left(1-\cos(2\xizeta)\right)
  {1\over2}\left(a_1^{\dagger}a_1 -  a_2^{\dagger}a_2\right) \nonumber \\
 &+&{{\bar \xi}^{2}\over2\xizeta^{2}}
 \left(-1+{\sin(2\xizeta)\over2\xizeta}\right)
 \left\{
 \cosh(2\zezeta)a_2^{\dagger}a_1 + 
 {{\bar \zeta}\sinh(2\zezeta)\over2\zezeta}\left(a_1\right)^2 +
 {\zeta\sinh(2\zezeta)\over2\zezeta}\left(a_2^{\dagger}\right)^2 
 \right\}, \\
 &&W^{-1}\frac{\partial}{\partial \zeta}W   \nonumber \\
 &=& 
 {1\over2}\left(1+{\sinh(2\zezeta)\over2\zezeta}\right)
  a_1^{\dagger}a_2^{\dagger} 
  + {{\bar \zeta}\over2\zezeta^{2}}\left(-1+ \cosh(2\zezeta)\right)
  {1\over2}\left(a_1^{\dagger}a_1 + a_2^{\dagger}a_2 + 1\right) \nonumber \\
 &+&{{\bar \zeta}^{2}\over2\zezeta^{2}}
 \left(-1+{\sinh(2\zezeta)\over2\xizeta}\right)a_1a_2.
\end{eqnarray}
From this lemma it is easy to calculate $A_{\xi}$ and $A_{\zeta}$. 
Let us here remember 
\[
   \ket{vac}=(\vert 0,0\rangle,\vert 0,1\rangle,
              \vert 1,0\rangle,\vert 1,1\rangle). 
\]

Before stating the result let us prepare some notations.
\begin{equation}
{\widehat E} = 
\left(
  \begin{array}{cccc}
    0& 0& 0& 0 \\
    0& 0& 1& 0 \\
    0& 0& 0& 0 \\
    0& 0& 0& 0 
  \end{array}
\right), 
{\widehat F} = 
\left(
  \begin{array}{cccc}
    0& 0& 0& 0 \\
    0& 0& 0& 0 \\
    0& 1& 0& 0 \\
    0& 0& 0& 0 
  \end{array}
\right), 
{\widehat H} = 
\left(
  \begin{array}{cccc}
    0& 0& 0& 0 \\
    0& {1\over2}& 0& 0 \\
    0& 0& -{1\over2}& 0 \\
    0& 0& 0& 0 
  \end{array}
\right).
\end{equation}
\begin{equation}
{\widehat A} = 
\left(
  \begin{array}{cccc}
    0& 0& 0& 1 \\
    0& 0& 0& 0 \\
    0& 0& 0& 0 \\
    0& 0& 0& 0 
  \end{array}
\right), 
{\widehat C} = 
\left(
  \begin{array}{cccc}
    0& 0& 0& 0 \\
    0& 0& 0& 0 \\
    0& 0& 0& 0 \\
    1& 0& 0& 0 
  \end{array}
\right), 
{\widehat B} = 
\left(
  \begin{array}{cccc}
    {1\over2}& 0& 0& 0 \\
    0& 1& 0& 0 \\
    0& 0& 1& 0 \\
    0& 0& 0& {3\over2}
  \end{array}
\right).
\end{equation}

\noindent{\bfseries Proposition 11}\quad We have
\begin{eqnarray}
 A_{\xi}&=& 
 {1\over2}\left(1+{\sin(2\xizeta)\over2\xizeta}\right)\cosh(2\zezeta)
 {\widehat F} 
  - {{\bar \xi}\over2\xizeta^{2}}\left(1-\cos(2\xizeta)\right){\widehat H} 
 \nonumber \\
  &&+ {{\bar \xi}^{2}\over2\xizeta^{2}}
 \left(-1+{\sin(2\xizeta)\over2\xizeta}\right)\cosh(2\zezeta){\widehat E}, \\
A_{\zeta} &=&  
 {1\over2}\left(1+{\sinh(2\zezeta)\over2\zezeta}\right){\widehat C}
  + 
 {{\bar \zeta}\over2\zezeta^{2}}\left(-1 + \cosh(2\zezeta)\right){\widehat B}
 \nonumber \\ 
  &&+ {{\bar \zeta}^{2}\over2\zezeta^{2}}
 \left(-1+{\sinh(2\zezeta)\over2\zezeta}\right){\widehat A}.
 \end{eqnarray}
Since the connection form $\cala$ is anti-hermitian 
($\cala^{\dagger}=-\cala$), 
it can be written as
\begin{equation}
%  \label{eq:calaa}
  \cala =
  A_\xi d\xi + A_\zeta d\zeta - A_\xi^{\dagger} d{\bar \xi}
  -A_\zeta^{\dagger} d{\bar \zeta}\ ,
\end{equation}
so that it's curvature form $\calf = d\cala + \cala\wedge\cala$ becomes
\begin{eqnarray}
  \label{eq:explicit}
  \calf
  &=&
  \left(
    \partial_\xi A_\zeta-\partial_\zeta A_\xi
    +[A_\xi,A_\zeta]
  \right) d\xi\wedge d\zeta
  \nonumber\\
  &&
  -\left(
    \partial_\xi A_\xi^\dagger+\partial_{\bar\xi}A_\xi
    +[A_\xi,A_\xi^\dagger]
  \right) d\xi\wedge d\bar\xi
  \nonumber\\
  &&
   -\left(
    \partial_\xi A_\zeta^\dagger+\partial_{\bar\zeta}A_\xi
    +[A_\xi,A_\zeta^\dagger]
  \right) d\xi\wedge d\bar\zeta
  \nonumber\\
  &&-\left(
    \partial_\zeta A_\xi^\dagger+\partial_{\bar\xi}A_\zeta
    +[A_\zeta,A_\xi^\dagger]
  \right) d\zeta\wedge d\bar\xi
  \nonumber\\
  &&-\left(
    \partial_\zeta A_\zeta^\dagger+\partial_{\bar\zeta}A_\zeta
    +[A_\zeta,A_\zeta^\dagger]
  \right) d\zeta\wedge d\bar\zeta
  \nonumber\\
  &&-\left(
    \partial_{\bar\xi}A_\zeta^\dagger-\partial_{\bar\zeta}A_\xi^\dagger
    +[A_\zeta^\dagger,A_\xi^\dagger]
  \right) d\bar\xi\wedge d\bar\zeta .
\end{eqnarray}

In this case we can calculate the curvature form completely. 
Now let us state our main result in this section.

\noindent{\bfseries Theorem 12}
\begin{eqnarray}
  \label{eq:main-result}
&& \calf = \nonumber\\
  &&
  -\Bigg\{
 \left(1+{\sin(2\xizeta)\over2\xizeta}\right)
 {{\bar \zeta}\sinh(2\zezeta)\over2\zezeta}  
 {\widehat F} + 
 {{\bar \xi}^{2}\over\xizeta^{2}}
 \left(-1+{\sin(2\xizeta)\over2\xizeta}\right)
 {{\bar \zeta}\sinh(2\zezeta)\over2\zezeta}  
 {\widehat E}
     \Bigg\}
  d\xi\wedge d\zeta
  \nonumber\\
  &&
  -\Bigg\{
  {\xi\over\xizeta^{2}}
  \left(-1+\cos(2\xizeta)\right)\cosh(2\zezeta){\widehat F}
  -{\sin(2\xizeta)\over\xizeta}
   \left(1+\cosh^2(2\zezeta)\right){\widehat H} \nonumber\\
 &&\ \ \ \ \ +{{\bar\xi}\over\xizeta^{2}}
 \left(-1+\cos(2\xizeta)\right)\cosh(2\zezeta){\widehat E}
     \Bigg\}
  d\xi\wedge d\bar\xi
  \nonumber\\
  &&
  -\Bigg\{
 \left(1+{\sin(2\xizeta)\over2\xizeta}\right)
 {\zeta\sinh(2\zezeta)\over2\zezeta}{\widehat F}  
 + 
 {{\bar\xi}^{2}\over\xizeta^{2}}
 \left(-1+{\sin(2\xizeta)\over2\xizeta}\right)
 {\zeta\sinh(2\zezeta)\over2\zezeta}{\widehat E}
     \Bigg\}
  d\xi\wedge d\bar\zeta
 \nonumber\\
  &&
  -\Bigg\{
 \left(1+{\sin(2\xizeta)\over2\xizeta}\right)
 {{\bar\zeta}\sinh(2\zezeta)\over2\zezeta}{\widehat E}  
 +
 {\xi^{2}\over\xizeta^{2}}
 \left(-1+{\sin(2\xizeta)\over2\xizeta}\right)
 {{\bar\zeta}\sinh(2\zezeta)\over2\zezeta}{\widehat F}
     \Bigg\}
  d\zeta\wedge d\bar\xi
 \nonumber\\
  &&
 - {\sinh(2\zezeta)\over\zezeta}\left(2{\widehat B}-{\textbf 1}_4\right)
  d\zeta\wedge d\bar\zeta
  \nonumber\\
  &&
  +\Bigg\{
 \left(1+{\sin(2\xizeta)\over2\xizeta}\right)
 {\zeta\sinh(2\zezeta)\over2\zezeta}{\widehat E}  
 +
 {\xi^{2}\over\xizeta^{2}}
 \left(-1+{\sin(2\xizeta)\over2\xizeta}\right)
 {\zeta\sinh(2\zezeta)\over2\zezeta}{\widehat F}
     \Bigg\}
  d\bar\xi\wedge d\bar\zeta.
\end{eqnarray}
From this and the theorem of Ambrose--Singer (see \cite{MN})  
it is easy to see that 

\noindent{\bfseries Corollary }
\begin{equation}
  Hol(\cala) = SU(2)\times U(1)\ \subset \ U(4).
\end{equation}
Therefore $\cala$ is not irreducible.

\subsection{Example $\cdots \ n=2$}

For this case we can also obtain the connection form in a complete manner. 
Let us perform.

\noindent{\bfseries Lemma 13}\quad We have
\begin{eqnarray}
 && W^{-1}\frac{\partial}{\partial \xis{1}}W = 
    {1\over2}\left(1+{\sin(2\xiszeta{1})\over2\xiszeta{1}}\right)
   \left\{
      \cos(\xiszeta{2})V^{-1}a_1^{\dagger}a_3V - 
   \frac{{\bar \xis{2}}\sin(\xiszeta{2})}{\xiszeta{2}}V^{-1}a_1^{\dagger}a_2V
   \right\}   \nonumber \\
  &+&
   {{\bar \xis{1}}\over2\xiszeta{1}^{2}}\left(1-\cos(2\xiszeta{1})\right)
   {1\over2}\left\{
     V^{-1}a_1^{\dagger}a_1V - 
     \frac{1+\cos(2\xiszeta{2})}{2} V^{-1}a_3^{\dagger}a_3V
   \right. \nonumber \\
   &+&
  \left.
  \frac{{\bar \xis{2}}\sin(2\xiszeta{2})}{2\xiszeta{2}}V^{-1}a_3^{\dagger}a_2V
   +
  \frac{\xis{2}\sin(2\xiszeta{2})}{2\xiszeta{2}}V^{-1}a_2^{\dagger}a_3V - 
   \frac{1-\cos(2\xiszeta{2})}{2} V^{-1}a_2^{\dagger}a_2V
    \right\}  \nonumber \\
  &+&
   {{\bar \xis{1}}^{2}\over2\xiszeta{1}^{2}}
       \left(-1+{\sin(2\xiszeta{1})\over2\xiszeta{1}}\right)
   \left\{
          \cos(\xiszeta{2})V^{-1}a_3^{\dagger}a_1V - 
   \frac{\xis{2}\sin(\xiszeta{2})}{\xiszeta{2}}V^{-1}a_2^{\dagger}a_1V
   \right\}, \\ 
  && \left. \right.  \nonumber \\ 
 && W^{-1}\frac{\partial}{\partial \xis{2}}W = 
    {1\over2}\left(1+{\sin(2\xiszeta{2})\over2\xiszeta{2}}\right)
    V^{-1}a_2^{\dagger}a_3V   \nonumber \\
  &+&  
   {{\bar \xis{2}}\over2\xiszeta{2}^{2}}\left(1-\cos(2\xiszeta{2})\right)
   {1\over2}\left(V^{-1}a_2^{\dagger}a_2V - V^{-1}a_3^{\dagger}a_3V \right)
    \nonumber \\
   &+& {{\bar \xis{2}}^{2}\over2\xiszeta{2}^{2}}
       \left(-1+{\sin(2\xiszeta{2})\over2\xiszeta{2}}\right)
   V^{-1}a_3^{\dagger}a_2V,   \\  
  && \left. \right.  \nonumber \\ 
 && W^{-1}\frac{\partial}{\partial \zetas{1}}W = 
    {1\over2}\left(1+{\sinh(2\zetaszeta{1})\over2\zetaszeta{1}}\right)
   \left\{
      \cosh(\zetaszeta{2})a_1^{\dagger}a_3^{\dagger} + 
   \frac{{\bar \zetas{2}}\sinh(\zetaszeta{2})}{\zetaszeta{2}}a_1^{\dagger}a_2
   \right\}   \nonumber \\
  &+&
 {{\bar \zetas{1}}\over2\zetaszeta{1}^{2}}\left(-1+\cosh(2\zetaszeta{1})\right)
  {1\over2}\left\{
   a_1^{\dagger}a_1 +   
     \frac{1+\cosh(2\zetaszeta{2})}{2} (a_3^{\dagger}a_3 + 1) + 
  \frac{{\bar \zetas{2}}\sinh(2\zetaszeta{2})}{2\zetaszeta{2}}a_2a_3 
   \right. \nonumber \\
   &+& 
   \left.
  \frac{\zetas{2}\sinh(2\zetaszeta{2})}{2\zetaszeta{2}}
   a_2^{\dagger}a_3^{\dagger} +  
   \frac{-1+\cosh(2\zetaszeta{2})}{2} a_2^{\dagger}a_2
    \right\}  \nonumber \\
  &+&
   {{\bar \zetas{1}}^{2}\over2\zetaszeta{1}^{2}}
       \left(-1+{\sinh(2\zetaszeta{1})\over2\zetaszeta{1}}\right)
   \left\{
          \cosh(\zetaszeta{2})a_1a_3 +  
   \frac{\zetas{2}\sinh(\zetaszeta{2})}{\zetaszeta{2}}a_2^{\dagger}a_1
   \right\},   \\  
  && \left. \right.  \nonumber \\ 
 && W^{-1}\frac{\partial}{\partial \zetas{2}}W =
    {1\over2}\left(1+{\sinh(2\zetaszeta{2})\over2\zetaszeta{2}}\right)
    a_2^{\dagger}a_3^{\dagger} + 
 {{\bar \zetas{2}}\over2\zetaszeta{2}^{2}}\left(-1+\cosh(2\zetaszeta{2})\right)
   {1\over2}\left(a_2^{\dagger}a_2 + a_3^{\dagger}a_3 + 1\right)
    \nonumber \\
   &+& {{\bar \zetas{2}}^{2}\over2\zetaszeta{2}^{2}}
       \left(-1+{\sinh(2\zetaszeta{2})\over2\zetaszeta{2}}\right)
    a_2a_3, 
\end{eqnarray}
where we remember $V = V_1V_2 = V_1(\zetas{1})V_2(\zetas{2}) =
{\rm exp}
{\left(\zetas{1} a_1^{\dagger}{a_3}^{\dagger} 
                    - {\bar \zetas{1}}a_{3}a_1 \right)}
{\rm exp}
{\left(\zetas{2} a_2^{\dagger}{a_3}^{\dagger} 
                    - {\bar \zetas{2}}a_{3}a_2 \right)}
$.

Next let us calculate $V^{-1}a_i^{\dagger}a_jV\ {\rm for}\ 1 \leq i,j \leq 3$.
From Lemma 9 we have

\noindent{\bfseries Corollary 14}
\begin{eqnarray}
V^{-1}a_1V &=& \cosh(\zetaszeta{1})a_1 +   
     \frac{\zetas{1}\sinh(\zetaszeta{1})}{\zetaszeta{1}}
   \left\{
     \cosh(\zetaszeta{2})a_3^{\dagger} + 
     \frac{{\bar \zetas{2}}\sinh(\zetaszeta{2})}{\zetaszeta{2}}a_2
   \right\}, \nonumber \\
V^{-1}a_2V &=& 
     \cosh(\zetaszeta{2})a_2 + 
     \frac{\zetas{2}\sinh(\zetaszeta{2})}{\zetaszeta{2}}a_3^{\dagger},
   \nonumber \\
V^{-1}a_3V &=& 
     \frac{\zetas{1}\sinh(\zetaszeta{1})}{\zetaszeta{1}}a_1^{\dagger} + 
   \cosh(\zetaszeta{1})
   \left\{
    \cosh(\zetaszeta{2})a_3 + 
     \frac{\zetas{2}\sinh(\zetaszeta{2})}{\zetaszeta{2}}a_2^{\dagger}
   \right\}. \nonumber 
\end{eqnarray}
From this we obtain

\noindent{\bfseries Lemma 15}
\begin{eqnarray}
&&V^{-1}a_1^{\dagger}a_1V = 
  \frac{1+\cosh(2\zetaszeta{1})}{2}a_1^{\dagger}a_1 +   
     \frac{\zetas{1}\sinh(2\zetaszeta{1})}{2\zetaszeta{1}}
  \left\{
    \cosh(\zetaszeta{2})a_1^{\dagger}a_3^{\dagger} + 
    \frac{{\bar \zetas{2}}\sinh(\zetaszeta{2})}{\zetaszeta{2}}a_1^{\dagger}a_2
  \right\} \nonumber \\
  &+&
     \frac{{\bar \zetas{1}}\sinh(2\zetaszeta{1})}{2\zetaszeta{1}}
  \left\{
    \cosh(\zetaszeta{2})a_3a_1 + 
    \frac{\zetas{2}\sinh(\zetaszeta{2})}{\zetaszeta{2}}a_2^{\dagger}a_1
  \right\} \nonumber \\
  &+&  
      \frac{-1+\cosh(2\zetaszeta{1})}{2}
  \left\{
   \frac{1+\cosh(2\zetaszeta{2})}{2}\left(a_3^{\dagger}a_3 + 1\right) + 
  \frac{{\bar \zetas{2}}\sinh(2\zetaszeta{2})}{2\zetaszeta{2}}a_3a_2
  \right. \nonumber \\
  &+&  \left. 
   \frac{\zetas{2}\sinh(2\zetaszeta{2})}{2\zetaszeta{2}}
   a_2^{\dagger}a_3^{\dagger} +
  \frac{-1+\cosh(2\zetaszeta{2})}{2}a_2^{\dagger}a_2
  \right\}, \nonumber \\
&& \left. \right. \nonumber \\
&&V^{-1}a_1^{\dagger}a_2V = 
    \cosh(\zetaszeta{1})
  \left\{
    \cosh(\zetaszeta{2})a_1^{\dagger}a_2 + 
  \frac{\zetas{2}\sinh(\zetaszeta{2})}{\zetaszeta{2}}a_1^{\dagger}a_3^{\dagger}
  \right\} \nonumber \\
  &+&
     \frac{{\bar \zetas{1}}\sinh(\zetaszeta{1})}{\zetaszeta{1}}
  \left\{
    \frac{1+\cosh(2\zetaszeta{2})}{2}a_3a_2 + 
 \frac{\zetas{2}\sinh(2\zetaszeta{2})}{2\zetaszeta{2}}
   \left(a_3^{\dagger}a_3 +1\right)
  \right\} \nonumber \\
  &+&
     \frac{\zetas{1}\sinh(\zetaszeta{1})}{\zetaszeta{1}}
  \left\{
 \frac{\zetas{2}\sinh(2\zetaszeta{2})}{2\zetaszeta{2}}a_2^{\dagger}a_2 + 
   \frac{\zetas{2}^2 \left(-1+\cosh(2\zetaszeta{2})\right)}{2\zetaszeta{2}^2}
  a_2^{\dagger}a_3^{\dagger}
  \right\}, \nonumber \\
&& \left. \right. \nonumber \\
&&V^{-1}a_1^{\dagger}a_3V = 
   \frac{\zetas{1}\sinh(2\zetaszeta{1})}{2\zetaszeta{1}}
      a_1^{\dagger}a_1^{\dagger} + 
   \cosh(2\zetaszeta{1})
  \left\{
    \cosh(\zetaszeta{2})a_1^{\dagger}a_3 + 
  \frac{\zetas{2}\sinh(\zetaszeta{2})}{\zetaszeta{2}}a_1^{\dagger}a_2^{\dagger}
  \right\} \nonumber \\
&+&
     \frac{{\bar \zetas{1}}\sinh(2\zetaszeta{1})}{2\zetaszeta{1}}
  \left\{
   \frac{1+\cosh(2\zetaszeta{2})}{2}a_3^2 + 
    \frac{2\sinh(2\zetaszeta{2})}{2\zetaszeta{2}}a_3a_2^{\dagger} +       
   \frac{\zetas{2}^2 \left(-1+\cosh(2\zetaszeta{2})\right)}{2\zetaszeta{2}^2}
  a_2^{\dagger}a_2^{\dagger}
  \right\}, \nonumber \\
&& \left. \right. \nonumber \\
&&V^{-1}a_2^{\dagger}a_2V =
      \frac{1+\cosh(2\zetaszeta{2})}{2} a_2^{\dagger}a_2 +   
  \frac{\zetas{2}\sinh(2\zetaszeta{2})}{2\zetaszeta{2}}
        a_2^{\dagger}a_3^{\dagger} + 
  \frac{-1+\cosh(2\zetaszeta{2})}{2}\left(a_3^{\dagger}a_3 + 1\right)
   \nonumber \\ 
&+& 
  \frac{{\bar \zetas{2}}\sinh(2\zetaszeta{2})}{2\zetaszeta{2}}a_3a_2 , 
   \nonumber \\
&& \left. \right. \nonumber \\
&&V^{-1}a_2^{\dagger}a_3V =
     \frac{\zetas{1}\sinh(\zetaszeta{1})}{\zetaszeta{1}}
  \left\{
      \cosh(\zetaszeta{2})a_2^{\dagger}a_1^{\dagger} + 
  \frac{{\bar \zetas{2}}\sinh(\zetaszeta{2})}{\zetaszeta{2}}a_3a_1^{\dagger}
  \right\} \nonumber \\
&+&
  \cosh(\zetaszeta{1})
  \left\{
   \frac{1+\cosh(2\zetaszeta{2})}{2}a_2^{\dagger}a_3 + 
    \frac{\zetas{2}\sinh(2\zetaszeta{2})}{2\zetaszeta{2}}
    a_2^{\dagger}a_2^{\dagger} +       
   \frac{{\bar \zetas{2}}\sinh(2\zetaszeta{2})}{2\zetaszeta{2}}a_3^2 + 
  \frac{-1+\cosh(2\zetaszeta{2})}{2} a_3a_2^{\dagger}
  \right\}, \nonumber \\
&& \left. \right. \nonumber \\
&&V^{-1}a_3^{\dagger}a_3V =
   \frac{-1+\cosh(2\zetaszeta{1})}{2}\left(a_1^{\dagger}a_1 +1\right) + 
\frac{{\bar \zetas{1}}\sinh(2\zetaszeta{1})}{2\zetaszeta{1}}
  \left\{
   \cosh(\zetaszeta{2})a_1a_3 +  
   \frac{\zetas{2}\sinh(\zetaszeta{2})}{\zetaszeta{2}}a_1a_2^{\dagger}
  \right\} \nonumber \\
&+&
\frac{\zetas{1}\sinh(2\zetaszeta{1})}{2\zetaszeta{1}}
  \left\{
   \cosh(\zetaszeta{2})a_3^{\dagger}a_1^{\dagger} +   
   \frac{{\bar \zetas{2}}\sinh(\zetaszeta{2})}{\zetaszeta{2}}a_2a_1^{\dagger}
  \right\} \nonumber \\
&+&
   \frac{1+\cosh(2\zetaszeta{1})}{2}
  \left\{
   \frac{1+\cosh(2\zetaszeta{2})}{2}a_3^{\dagger}a_3 + 
    \frac{\zetas{2}\sinh(2\zetaszeta{2})}{2\zetaszeta{2}}
    a_3^{\dagger}a_2^{\dagger} + 
   \frac{{\bar \zetas{2}}\sinh(2\zetaszeta{2})}{2\zetaszeta{2}}a_2a_3 
\right. \nonumber \\
&+&
\left.  
  \frac{-1+\cosh(2\zetaszeta{2})}{2}\left(a_2^{\dagger}a_2 + 1\right)
  \right\}. \nonumber 
 \end{eqnarray}
Making use of this lemma it is not difficult to calculate $A_{\xis{j}}$ and 
$A_{\zetas{j}}\ (j=1, 2)$. 
Let us again remember 
\[
   \ket{vac}=(\vert 0,0,0\rangle,\vert 0,0,1\rangle,
              \vert 0,1,0\rangle,\vert 0,1,1\rangle,
              \vert 1,0,0\rangle,\vert 1,0,1\rangle,
              \vert 1,1,0\rangle,\vert 1,1,1\rangle
              ). 
\]
Therefore we have only to know that for $1 \leq i, j \leq 3$
\[
  \bra{vac}a_i^{\dagger}a_j\ket{vac},\  \bra{vac}a_ia_j\ket{vac}\ {\rm and}\  
  \bra{vac}V^{-1}a_i^{\dagger}a_jV\ket{vac}.
\]
First let us determine $\bra{vac}a_i^{\dagger}a_j\ket{vac}$ and $
\bra{vac}a_ia_j\ket{vac}$.

\noindent{\bfseries Lemma 16}
\[
\bra{vac}a_1^{\dagger}a_1\ket{vac}= 
\left(
  \begin{array}{cccccccc}
    0& 0& 0& 0& 0& 0& 0& 0 \\
    0& 0& 0& 0& 0& 0& 0& 0 \\
    0& 0& 0& 0& 0& 0& 0& 0 \\
    0& 0& 0& 0& 0& 0& 0& 0 \\
    0& 0& 0& 0& 1& 0& 0& 0 \\
    0& 0& 0& 0& 0& 1& 0& 0 \\
    0& 0& 0& 0& 0& 0& 1& 0 \\
    0& 0& 0& 0& 0& 0& 0& 1 
   \end{array}
\right),
\bra{vac}a_1^{\dagger}a_2\ket{vac}= 
\left(
  \begin{array}{cccccccc}
    0& 0& 0& 0& 0& 0& 0& 0 \\
    0& 0& 0& 0& 0& 0& 0& 0 \\
    0& 0& 0& 0& 0& 0& 0& 0 \\
    0& 0& 0& 0& 0& 0& 0& 0 \\
    0& 0& 1& 0& 0& 0& 0& 0 \\
    0& 0& 0& 1& 0& 0& 0& 0 \\
    0& 0& 0& 0& 0& 0& 0& 0 \\
    0& 0& 0& 0& 0& 0& 0& 0 
   \end{array}
\right), \nonumber
\]
\[
\bra{vac}a_1^{\dagger}a_3\ket{vac}= 
\left(
  \begin{array}{cccccccc}
    0& 0& 0& 0& 0& 0& 0& 0 \\
    0& 0& 0& 0& 0& 0& 0& 0 \\
    0& 0& 0& 0& 0& 0& 0& 0 \\
    0& 0& 0& 0& 0& 0& 0& 0 \\
    0& 1& 0& 0& 0& 0& 0& 0 \\
    0& 0& 0& 0& 0& 0& 0& 0 \\
    0& 0& 0& 1& 0& 0& 0& 0 \\
    0& 0& 0& 0& 0& 0& 0& 0 
   \end{array} 
\right),  
\bra{vac}a_1a_2\ket{vac}= 
\left(
  \begin{array}{cccccccc}
    0& 0& 0& 0& 0& 0& 1& 0 \\
    0& 0& 0& 0& 0& 0& 0& 1 \\
    0& 0& 0& 0& 0& 0& 0& 0 \\
    0& 0& 0& 0& 0& 0& 0& 0 \\
    0& 0& 0& 0& 0& 0& 0& 0 \\
    0& 0& 0& 0& 0& 0& 0& 0 \\
    0& 0& 0& 0& 0& 0& 0& 0 \\
    0& 0& 0& 0& 0& 0& 0& 0 
   \end{array} 
\right), \nonumber 
\]
\[
\bra{vac}a_1a_3\ket{vac}= 
\left(
  \begin{array}{cccccccc}
    0& 0& 0& 0& 0& 1& 0& 0 \\
    0& 0& 0& 0& 0& 0& 0& 0 \\
    0& 0& 0& 0& 0& 0& 0& 1 \\
    0& 0& 0& 0& 0& 0& 0& 0 \\
    0& 0& 0& 0& 0& 0& 0& 0 \\
    0& 0& 0& 0& 0& 0& 0& 0 \\
    0& 0& 0& 0& 0& 0& 0& 0 \\
    0& 0& 0& 0& 0& 0& 0& 0 
   \end{array} 
\right),  
\bra{vac}a_2^{\dagger}a_2\ket{vac}= 
\left(
  \begin{array}{cccccccc}
    0& 0& 0& 0& 0& 0& 0& 0 \\
    0& 0& 0& 0& 0& 0& 0& 0 \\
    0& 0& 1& 0& 0& 0& 0& 0 \\
    0& 0& 0& 1& 0& 0& 0& 0 \\
    0& 0& 0& 0& 0& 0& 0& 0 \\
    0& 0& 0& 0& 0& 0& 0& 0 \\
    0& 0& 0& 0& 0& 0& 1& 0 \\
    0& 0& 0& 0& 0& 0& 0& 1 
   \end{array} 
\right),  
\]
\[
\bra{vac}a_2^{\dagger}a_3\ket{vac}= 
\left(
  \begin{array}{cccccccc}
    0& 0& 0& 0& 0& 0& 0& 0 \\
    0& 0& 0& 0& 0& 0& 0& 0 \\
    0& 1& 0& 0& 0& 0& 0& 0 \\
    0& 0& 0& 0& 0& 0& 0& 0 \\
    0& 0& 0& 0& 0& 0& 0& 0 \\
    0& 0& 0& 0& 0& 0& 0& 0 \\
    0& 0& 0& 0& 0& 1& 0& 0 \\
    0& 0& 0& 0& 0& 0& 0& 0 
   \end{array} 
\right),  
\bra{vac}a_2a_3\ket{vac}= 
\left(
  \begin{array}{cccccccc}
    0& 0& 0& 1& 0& 0& 0& 0 \\
    0& 0& 0& 0& 0& 0& 0& 0 \\
    0& 0& 0& 0& 0& 0& 0& 0 \\
    0& 0& 0& 0& 0& 0& 0& 0 \\
    0& 0& 0& 0& 0& 0& 0& 1 \\
    0& 0& 0& 0& 0& 0& 0& 0 \\
    0& 0& 0& 0& 0& 0& 0& 0 \\
    0& 0& 0& 0& 0& 0& 0& 0 
   \end{array} 
\right),  
\]
\[
\bra{vac}a_3^{\dagger}a_3\ket{vac}= 
\left(
  \begin{array}{cccccccc}
    0& 0& 0& 0& 0& 0& 0& 0 \\
    0& 1& 0& 0& 0& 0& 0& 0 \\
    0& 0& 0& 0& 0& 0& 0& 0 \\
    0& 0& 0& 1& 0& 0& 0& 0 \\
    0& 0& 0& 0& 0& 0& 0& 0 \\
    0& 0& 0& 0& 0& 1& 0& 0 \\
    0& 0& 0& 0& 0& 0& 0& 0 \\
    0& 0& 0& 0& 0& 0& 0& 1 
   \end{array} 
\right), \ \quad 
   {\rm and}\qquad  \bra{vac}a_j^{2}\ket{vac}= {\mathbf 0}_8 \quad j=1,2,3.
\] 

Next let us determine $\bra{vac}V^{-1}a_i^{\dagger}a_jV\ket{vac}$. For that 
we prepare some notations : for $1 \leq i, j \leq 3$ we set 
\[
   M_{ij}= \bra{vac}a_i^{\dagger}a_j\ket{vac},\quad 
   N_{ij}= \bra{vac}a_ia_j\ket{vac}.
\]
Then both $M_{ij}$ and $N_{ij}$ are real matrices and moreover satisfy 
\begin{eqnarray}
 &&M_{ij}^{\dagger}=M_{ji} \quad {\rm and}\quad 
  M_{ii}\ne {\mathbf 0}_8, \nonumber \\
 &&N_{ij}=N_{ji}\quad {\rm and}\quad N_{ii}={\mathbf 0}_8. \nonumber
\end{eqnarray}
Then we have 

\noindent{\bfseries Lemma 17}
\begin{eqnarray}
&&\bra{vac}V^{-1}a_1^{\dagger}a_1V\ket{vac}=
    \frac{1+\cosh(2\zetaszeta{1})}{2}M_{11} + 
    \frac{\zetas{1}\sinh(2\zetaszeta{1})}{2\zetaszeta{1}}
  \left\{
    \cosh(\zetaszeta{2})N_{13}^{\dagger} + 
    \frac{{\bar \zetas{2}}\sinh(\zetaszeta{2})}{\zetaszeta{2}}M_{12}
  \right\} \nonumber \\
&+&
    \frac{{\bar \zetas{1}}\sinh(2\zetaszeta{1})}{2\zetaszeta{1}}
  \left\{
    \cosh(\zetaszeta{2})N_{13} + 
    \frac{\zetas{2}\sinh(\zetaszeta{2})}{\zetaszeta{2}}M_{12}^{\dagger}
  \right\} + 
  \frac{-1+\cosh(2\zetaszeta{1})}{2}
  \left\{
   \frac{1+\cosh(2\zetaszeta{2})}{2} \left(M_{33}+E \right) 
\right. \nonumber \\
&+& 
\left.
  \frac{{\bar \zetas{2}}\sinh(2\zetaszeta{2})}{2\zetaszeta{2}}N_{23} + 
   \frac{\zetas{2}\sinh(2\zetaszeta{2})}{2\zetaszeta{2}}N_{23}^{\dagger} +
  \frac{-1+\cosh(2\zetaszeta{2})}{2}M_{22}
  \right\}, \\
&& \left. \right. \nonumber \\
&&\bra{vac}V^{-1}a_1^{\dagger}a_2V\ket{vac}=
    \cosh(\zetaszeta{1})
  \left\{
    \cosh(\zetaszeta{2})M_{12} + 
  \frac{\zetas{2}\sinh(\zetaszeta{2})}{\zetaszeta{2}}N_{13}^{\dagger}
  \right\} \nonumber \\
  &+&
     \frac{{\bar \zetas{1}}\sinh(\zetaszeta{1})}{\zetaszeta{1}}
  \left\{
    \frac{1+\cosh(2\zetaszeta{2})}{2}N_{23} + 
 \frac{\zetas{2}\sinh(2\zetaszeta{2})}{2\zetaszeta{2}}
   \left(M_{33}+E\right)
  \right\} \nonumber \\
  &+&
     \frac{\zetas{1}\sinh(\zetaszeta{1})}{\zetaszeta{1}}
  \left\{
 \frac{\zetas{2}\sinh(2\zetaszeta{2})}{2\zetaszeta{2}}M_{22} + 
   \frac{\zetas{2}^2 \left(-1+\cosh(2\zetaszeta{2})\right)}{2\zetaszeta{2}^2}
  N_{23}^{\dagger}
  \right\}, \\
&& \left. \right. \nonumber \\
&&V^{-1}a_1^{\dagger}a_3V = 
   \cosh(2\zetaszeta{1})
  \left\{
    \cosh(\zetaszeta{2})M_{13} + 
  \frac{\zetas{2}\sinh(\zetaszeta{2})}{\zetaszeta{2}}N_{12}^{\dagger}
  \right\} \nonumber \\
&+&
    \frac{{\bar \zetas{1}}\sinh(2\zetaszeta{1})}{2\zetaszeta{1}}
    \frac{2\sinh(2\zetaszeta{2})}{2\zetaszeta{2}}M_{23}, \\
&& \left. \right. \nonumber \\ 
&&V^{-1}a_2^{\dagger}a_2V =
      \frac{1+\cosh(2\zetaszeta{2})}{2} M_{22} +   
  \frac{\zetas{2}\sinh(2\zetaszeta{2})}{2\zetaszeta{2}}N_{23}^{\dagger} + 
  \frac{-1+\cosh(2\zetaszeta{2})}{2}\left(M_{33} + E\right)
   \nonumber \\ 
&+& 
  \frac{{\bar \zetas{2}}\sinh(2\zetaszeta{2})}{2\zetaszeta{2}}N_{23} , \\
&& \left. \right. \nonumber \\
&&V^{-1}a_2^{\dagger}a_3V =
     \frac{\zetas{1}\sinh(\zetaszeta{1})}{\zetaszeta{1}}
  \left\{
      \cosh(\zetaszeta{2})N_{12}^{\dagger} + 
  \frac{{\bar \zetas{2}}\sinh(\zetaszeta{2})}{\zetaszeta{2}}M_{13}
  \right\} 
+
  \cosh(\zetaszeta{1})\cosh(2\zetaszeta{2})M_{23}, \\
&& \left. \right. \nonumber \\
&&V^{-1}a_3^{\dagger}a_3V =
   \frac{-1+\cosh(2\zetaszeta{1})}{2}\left(M_{11} +E\right) + 
\frac{{\bar \zetas{1}}\sinh(2\zetaszeta{1})}{2\zetaszeta{1}}
  \left\{
   \cosh(\zetaszeta{2})N_{13} +  
   \frac{\zetas{2}\sinh(\zetaszeta{2})}{\zetaszeta{2}}M_{12}^{\dagger}
  \right\} \nonumber \\
&+&
\frac{\zetas{1}\sinh(2\zetaszeta{1})}{2\zetaszeta{1}}
  \left\{
   \cosh(\zetaszeta{2})N_{13}^{\dagger} +   
   \frac{{\bar \zetas{2}}\sinh(\zetaszeta{2})}{\zetaszeta{2}}M_{12}
  \right\} \nonumber \\
&+&
   \frac{1+\cosh(2\zetaszeta{1})}{2}
  \left\{
   \frac{1+\cosh(2\zetaszeta{2})}{2}M_{33} + 
    \frac{\zetas{2}\sinh(2\zetaszeta{2})}{2\zetaszeta{2}}
    N_{23}^{\dagger} + 
   \frac{{\bar \zetas{2}}\sinh(2\zetaszeta{2})}{2\zetaszeta{2}}N_{23}
\right. \nonumber \\
&+&
\left.  
  \frac{-1+\cosh(2\zetaszeta{2})}{2}\left(M_{22} + E\right)
  \right\}.  
\end{eqnarray}
Here we have denoted by $E$ the unit matrix in $M(8,\fukuso)$.

Using these lemmas we can obtain $A_{\xis{j}}$ and 
$A_{\zetas{j}}\ (j=1, 2)$ completely. Next we must calculate the 
curvature form making use of the connection form, but it is too 
hard. We leave its calculation to interested readers.

\section{Discussion}

We in this paper defined unitary coherent operators based on Lie algebras
$su(n+1)$ and $su(n,1)$ and, making use of these, 
calculated non--abelian Berry connections 
of quantum computational bundles proposed by Zanardi and Rasetti \cite{ZR}
.  For $n = 1$ and $2$ we gave an explicit form to them.
This ia a generalization of that of Pachos and Chountasis \cite{PC}.

But for $n \geq 3$ we could not give explicit ones due to complexity. 
Therefore our paper is far from complete. 
As $n$ becomes large our culculation will become miserable. Moreover 
we didn't perform the calculation of curvatures except for $n=1$.

We have a lot of problems to be performed. We expect that many young 
mathematical physicists with brute force will enter in this field.

\noindent{\em Acknowledgment.}\\
The author wishes to thank K. Funahashi and Y. Machida for their helpful 
comments and suggestions. 
I also thank J. Pachos and S. Chountasis for some useful suggestions.

%%%%%%%%%%%%
%References%
%%%%%%%%%%%%

\end{document}